 \documentclass[twocolumn,nofootinbib,10pt]{revtex4-1}
\usepackage{graphicx,float}\usepackage[all]{xy}
\usepackage{amsmath,upgreek}
\usepackage{amssymb}
\usepackage{color}
\usepackage[colorlinks,hyperindex]{hyperref}

\newcommand{\bes}{\begin{subequations}}
\newcommand{\ees}{\end{subequations}}
\def\ben{\begin{eqnarray}}
\def\een{\end{eqnarray}}
\def\be{\begin{equation}}
\def\ee{\end{equation}}

\begin{document}

\title { Information versus stability in an anti-de Sitter black hole }
 \author{Nelson R. F. Braga}
\affiliation{Instituto de F\'isica, Universidade Federal do Rio de Janeiro,
Caixa Postal 68528, RJ 21941-972 - Brazil}
\email{braga@if.ufrj.br}


\begin{abstract}

Information entropies associated with the energy density  in position and momentum spaces  are build for an anti-de Sitter (AdS)  black hole. These quantities, that satisfy an entropic uncertainty relation,
vary with the temperature. The higher is  the black hole temperature, the greater/smaller is the information encoded respectively in the position/momentum  distributions of energy. 
On the other hand, as it is well know, AdS black holes are subject to the Hawking Page phase transition. The amplitude for dominance of the black hole phase over the thermal AdS phase increases with the temperature.   So, as the system becomes more stable, there is a change in the way  that   information is stored.  In particular:  information stored  in the spatial energy density increases while information stored in the energy density in momentum space decreases.

\end{abstract}

\pacs{ }
\keywords{Information  entropy,  Anti-de sitter black hole, Quantum Information}

\maketitle    
 
 \section{ Introduction}
 
 It has been proposed by Gleiser and Sowinski \cite{Gleiser:2013mga} that the stability of compact objects can be characterized by a quantity,  called  configuration entropy,  introduced by  Gleiser and Stamatopoulos in refs.\cite{Gleiser:2011di,Gleiser:2012tu}.  The smaller is this quantity, the more stable is the system. In the recent years, many different examples appeared in the literature where there is a similar relation between the variation of the  configurational entropy and the stability was shown to appear in as diverse systems as compact astrophysical objects    
\cite{Gleiser:2015rwa}  and holographic AdS/QCD models \cite{Bernardini:2016hvx,Bernardini:2016qit,Braga:2017fsb,Braga:2018fyc,Bernardini:2018uuy,Ferreira:2019inu}. Many other recent applications of configuration entropy are found in the literature, as for example
 \cite{Alves:2014ksa,Gleiser:2015rwa,Correa:2015vka,Sowinski:2017hdw,Casadio:2016aum,Correa:2016pgr,Braga:2016wzx,Karapetyan:2016fai,Karapetyan:2017edu,Alves:2017ljt,Karapetyan:2018oye,Karapetyan:2018yhm,Gleiser:2018kbq,Lee:2018zmp,Bazeia:2018uyg,Ma:2018wtw,Zhao:2019xle}

The basis for the definition of the configuration entropy\cite{Gleiser:2013mga}  is the 
 information entropy of Shannon \cite{shannon} that has the form
 \begin{equation}
  - \sum_n  \, p_n  \log p_n \,,
 \label{discretepositionentropy}
 \end{equation}
 where $p_n$ is the probability distribution of a discrete variable. For the configuration entropy, one considers the continous version 
  \begin{equation}
 f = - \int d^3r \, \rho({\vec r}) \log \rho({\vec r}) \,,
 \label{positionentropy}
 \end{equation}
 where $\rho ({\vec r}) = \vert v ({\vec r}) \vert^2  $ is a normalized funtion $\int d^3r \rho({ \vec r }) =1$, called modal fraction. It is important to note the discrete case of eq. (\ref{discretepositionentropy})
 is positive definite since the probabilities satisfies $p_n \le 1$. The same rule does not apply to the continuous case, that involves densities. So, one should not take eq. (\ref{positionentropy}) as an absolute measure of the information content, but rather consider the variations of $f$ as representing variation in the information content. 
  
  The configuration entropy, as defined in \cite{Gleiser:2013mga},  is the momentum space version of $f$
 \begin{equation}
 {\tilde f } = - \int d^3k \, { \tilde \rho} ({\vec k}) \log  {\tilde \rho} ({\vec k}) \,,
 \label{momentumentropy}
 \end{equation}
 where ${\tilde \rho } ({\vec k}) = \vert {\tilde v } ({\vec k}) \vert^2  $ is the momentum space modal fraction, with
  \begin{equation}
 {\tilde v } (\vec k )  = \frac{1}{(2\pi)^{3/2}} \int d^3r \,  v ({\vec x}) \exp ( -i \vec k \cdot \vec r )   \,. 
 \label{Conjugate}
 \end{equation}
 Information entropies like $f$ and ${\tilde f} $, based in conjugate variables in the sense of eq. (\ref{Conjugate}) satisfy the so called entropic uncertainty relations \cite{Birula}, that, for this 3-dimensional case takes the form:
 \begin{equation}
 f + {\tilde f}  \, \ge 3(1 + \log \pi ) \,.
 \label{uncertainty}
  \end{equation} 
  
  So, one could guess that a variation of the configuration entropy, defined in momentum space,  ${\tilde f} $ should be associated with a variation of the conjugate quantity, $ f$, defined in  position space. 
The purpose of this letter is to go one step ahead and investigate the  relation between momentum and position entropies and their relation with stability. 
 
The idea is to use the anti-de Sitter (AdS)  black hole as an  example,  motivated by the fact that for this physical system one finds a simple way to characterize what one means by stability. That is:  stability of the black hole is related to the amplitude for the  dominance of the black hole phase over the thermal AdS phase, as we will see in section II. 
   
 A discussion of the configuration entropy  for an anti-de Sitter black hole already appeared in ref.  \cite{Braga:2016wzx}.  However, there two completely new aspects in the present letter. One is a technical point. That is: in ref. \cite{Braga:2016wzx}  the energy density was obtained by using a particular regularization in order to get rid of a surface contribution to the mass.  This regularization process, as we will discuss in section III, is not unique. So, the definition for the energy density was ambiguous. 
 We will present in section III an alternative  non ambiguous way of introducing the energy density.

The second point is that, in contrast to ref. \cite{Braga:2016wzx}, here we will investigate not only the configuration entropy  in momentum space ${\tilde f} $ but also the corresponding dual entropy  $f$ in position space.  We will see that  these two quantities, that are subject to the inequality (\ref{uncertainty}),  
 vary with the black hole temperature. However, the sum ${\tilde f} + f$ 
is constant for the black hole case. So that, if these two quantities represent the information content in momentum and position spaces, respectively, such a result indicates that the total information is conserved. 

 There is a non trivial point regarding the interpretation of the entropies that we will calculated using the black hole energy densities.  Information is associated  with the degree of unpredictability  of the result of an observation. In the Shannon entropy of eq. (\ref{discretepositionentropy}) the factors 
 $p_i$ are  probabilities. As a simple  illustration of the relation between the information content and the degree of unpredictability, one can consider the particular case of the Shannon entropy when there  is just one possible result for an observation. In this situation, there is just one value of $i$, with $p_i = 1 $ and the entropy vanishes. In simple words: when there is no unpredictability the information entropy vanishes. 
 
 For the continuum case, in place of the discrete probability, one uses the modal fractions $  \rho  ({\vec x}) $   and ${ \tilde \rho} ({\vec k})$  in the definitions of the  entropies in eqs.  (\ref{positionentropy}) and 
(\ref{momentumentropy}) respectively.  If it happens that the modal fractions can be interpreted as probability densities, then the entropies      $f$ and  ${\tilde f} $ can be interpreted as representing the information content in position and momentum spaces, respectively. In this letter we will build up modal fractions as the normalized square of the energy densities of the AdS black hole, in  position and momentum spaces. These quantities are not probability densities if one makes an observation of the energy. The energy distribution is completely known, once the density is given. So, regarding energy distribution there is no unpredictability.  However, one can think about an  ``experiment'' where one  searches for the position of a particle and the probability density of finding the particle at a given position is  the normalized square of the spatial energy density. In a similar way, for the momentum case one can consider a process of determining the momentum of a particle that has a probability  density in momentum space equal to the modal fraction obtained from the energy density in momentum space.
It is in this sense that we will 
 interpret  $f$ and ${\tilde f} $ as representing information content. 
 
 This letter is organized as follows.  In section II we briefly review the Hawking Page transition in an AdS black hole. In section II we present the new approach for finding the energy density of the AdS black hole. Then in section IV we present the results for the entropies  and  some final conclusions.

\section{ Hawking Page transition }    
A very interesting description of an anti-de Sitter (AdS)  black hole was found by Hawking and Page \cite{Hawking:1982dh}  (see also \cite{Witten:1998zw} for enlightening discussions). 
The point of view, following semi-classical arguments, is that the black hole is a physical system consisting of a superposition of two different geometries with the same asymptotic boundaries. Both are solutions of  the vacuum Einstein equation with a negative cosmological constant.  One is the  (thermal)  Euclidean AdS space
\begin{equation}
\label{AdS}
 ds^2 \,\,= \,\, \left(1+\frac{r^2}{b^2}\right) \,d{\bar t}^{\, 2 }+ \frac{dr^2}{1+\frac{r^2}{b^2}} + r^2 d\Omega^2_{(2)} \,\,,
 \end{equation}
with a boundary, at $r \to \infty$, that is the product of  a spatial ${\boldmath S}^{2}$ sphere with a temporal  $S^1$ circle. The time coordinate is periodic:  $ {\bar t} \sim  {\bar t} +\bar \beta$.  
 
 The other geometry is the  AdS-Schwarzschild black hole space 
 \begin{equation}
 \label{AdSBH}
 ds^2 \,\,= \, \left(1+\frac{r^2}{b^2} - \frac{ 2M G_N}{r}\right) \,dt^2 +\frac{dr^2}{1+\frac{r^2}{b^2}- \frac{ 2M G_N}{r} }      + r^2 d\Omega^2 \,
 \end{equation}
 where  $M$ is the black hole mass and  $G_N $ is the Newton constant. In this case the temporal periodicity is $ t \sim  t +\beta$.  The black hole space is the region $ r > r_h$, were  $ r_h$, for our purposes,  is the largest root of: 
  \begin{equation}
 \label{horizon}
1+ \frac{r^2}{b^2}  \,- \frac{2 M G_N}{r } \,= 0 \,.
  \end{equation} 
 In order to avoid a conical singularity, the temporal period must be  
 \begin{equation}
 \label{tempXhorizon}
\beta =  \frac{4 \pi b^2 r_h }{3 r_h^2 +  b^2}   \,.
  \end{equation}
The black hole temperature is $T = 1/ \beta.$ 

The  Einstein gravity  actions for  the geometries (\ref{AdS}) and (\ref{AdSBH}) are respectively
 \begin{eqnarray} 
  I_{\rm AdS} &=&   \frac{3}{ 2 \, b^2 G_N}   \, \int_0^{\bar \beta} d\bar{t}  \int_{0}^R r^{2} dr \,\,,
  \label{THAdSAction}\\
  I_{\rm BH} &=&  \frac{3}{ 2 \, b^2 G_N} \, \int_0^{\beta} dt \int_{r_h}^R r^{2} dr  \,\,,
  \label{BHaction}  
    \end{eqnarray}
 where $R$ is a large (regulator) radius.  
   In order that  the actions have the same asymptotic geometry at $r\to \infty$, one  imposes   that the  temporal circles have equal length: 
\begin{equation}
 \label{relationoftimes}
\sqrt{ \left(1+ \frac{R^2}{b^2} - \frac{ 2 M G_N}{R} \right) }\,\,\,dt \,=\,  \sqrt{ \left(1+ \frac{R^2}{b^2}\right) } \,\,\,d\bar{t} \,\,.
 \end{equation}
 This implies that at large  $R $:  
 \begin{equation}
 \label{relationoperiods}
 {\bar \beta} \,\approx \, \beta \, \sqrt{ 1 \,-\, \frac{ r_h^3 + b^2 r_h  }{ 2 R^{3}  }}
  \,\,.
 \end{equation}
  
 The regularized action $I$, that represents the effect of the presence of the black hole, with respect to the AdS space without the black hole, reads
\begin{equation} 
I  = I_{\rm BH} \,-\,  I_{\rm AdS}  \,,
  \label{Regularized}
 \end{equation}    
 and is finite in the $ R \to \infty $ limit. 
  
 The black hole solutions considered here are restricted to  $ b/\sqrt{3} <  r_h  $. The action $I$ is positive for $ r_h <  b$, when the AdS space is dominant and is negative for  $ r_h > b$, when the black hole space become  dominant.  The amplitude for finding
 the black hole geometry is approximately governed by the factor $\exp \left(\, - I  \, \right)$.
Since the action $I $ decreases monotonically with $r_h$, the black hole becomes more and more stable against the Hawking Page transition as $r_h$ and, correspondingly  the temperature $T$, increase.

  \section{Energy density for the AdS black hole}
  
  The mass of the  AdS black hole is obtained from  \cite{Witten:1998zw} 
    \begin{equation}
 \label{TotalMass}
  M = \frac{\partial I}{\partial \beta} \,,
  \end{equation}
  with the action $I $ given by eqs.  (\ref{THAdSAction}) ,  (\ref{BHaction}) and (\ref{Regularized}). The idea that we will follow, in order to find an energy density, is to find a way of writing the total mass of eq. (\ref{TotalMass}) as a spatial integral:  
  \begin{equation}
 \label{ massfromdensity3d}
M =  \int u ( r )  \, d^3r    \,. 
\end{equation}
This means that we have to commute the derivative with respect to $\beta $ with the spatial integration, contained in the action integral $I$.  
This is not a trivial task since there is subtle point in equation (\ref{TotalMass}). The limit of the radial integration of the black hole action depends on the temperature. This happens because the black hole space is defined as the region outside the horizon, as can be seen in eq. (\ref{BHaction}) and the horizon position depends on the temperature, as show in eq. (\ref{tempXhorizon}). 
So, the derivative with respect to $\beta (= 1/T) $ affects not only the temporal periods but also the spatial part of the action integral. In ref. \cite{Braga:2016wzx}, a proposal to define an energy density was presented. But the approach used there was to use the standard radial coordinate, as in eq. (\ref{BHaction}), where the dependence on the temperature appears in the lower integration limit. Then, the derivative with respect to $\beta $ lead to a surface term to the mass. This surface term comes from the derivative of the radial integration limit and in ref. \cite{Braga:2016wzx} it was located at the horizon position. This  term would give a singular contribution to the entropy, as a consequence of the singular localization in the radial coordinate. In order to fix this problem, it was proposed in \cite{Braga:2016wzx} that the surface term should be replaced by a volume density that leads to the same contribution to the mass.  This was interpreted as a regularization procedure. However,  it was not possible  to find a unique definition for the volume density. So,  a consistent definition for the energy density was lacking. 

Here we will propose a different approach where one does  not need to introduce any arbitrary term in the density. We simply perform a change of the radial variable to $ x = r / r_h $ that moves the dependence on the temperature from the lower radial integration limit to the upper integration limit. The actions take the form 
  \begin{eqnarray} 
  I_{\rm AdS} &=&   \frac{3}{ 2 \, b^2 G_N}   \, {\bar \beta} \,  r_h^3  \int_{0}^{R/r_h} x^{2} dx \,\,,
  \label{newTHAdSAction}\\I_{\rm BH} &=&  \frac{3}{ 2 \, b^2 G_N} \, \beta \, r_h^3 \int_{1}^{R/r_h} x^{2} dx \,\,.
  \label{newBHaction}  
    \end{eqnarray}

  Differentiating $ I  = I_{\rm BH} \,-\,  I_{\rm AdS}$  with respect to $\beta$, using  relations 
  (\ref{tempXhorizon}) and (\ref{relationoperiods})  and then taking the limit $ R \to \infty$, one finds the mass of the black hole expressed as  just a  volume integral. Surface terms appear when one differentiates with respect to the upper integration limit $ x = R/r_h $. However, in contrast to what happens in the approach of ref. \cite{Braga:2016wzx}, they vanish when one subtracts the contributions from the black hole and thermal AdS actions and then take the $ R \to \infty$ limit. 
  
  One finds two different constant densities: $u_1(x) $  in the region $ 0 \le x \le 1 $ and $u_2 (x) $   for  the region 
  $1 \le x \le R/r_h$. In other words, the mass can be written as
  \begin{equation}
 \label{ massfromdensity}
M =  4 \pi \int_{0}^{R/r_h} x^2  u (x)    dx    \,, 
\end{equation}
with
\begin{equation} 
u(x) =\begin{cases}\frac{3}{8 \pi b^2 G_N} \left( \frac{6 r_h^5 + 4 r_h^3 b^2 }{3 r_h^2 - b^2} \right), &  (0 \le x \le 1)\,,  \\  \\ - \frac{3  r_h^3 }{8 \pi   b^2 G_N} \left( \frac{3 r_h^5 +2  r_h^3 b^2 + r_h b^4}{   R^3 \, ( 3 r_h^2 - b^2 ) } \right),&(1 \le x \le \frac{R}{r_h}) \, . \end{cases}
\label{densities}
\end{equation}
\noindent Note that the energy density is proportional to $ 1/R^3 $. In the $R \to \infty $  limit the density goes to zero, but the contribution to the mass is finite since the volume increases with $R^3$.  
The density in the region $r > r_h$, outside the horizon, will not contribute to the information entropies. 

Now we return to coordinate $r$ and define the corresponding mass/energy density:
$$ v (r) = u (x) /r_h^3 \, .$$ 
The modal fraction that we need, in order to define the information entropy in position space  is 
\begin{equation} 
\rho (r) = \frac{ \vert v(r) \vert^2 }{ 4 \pi \int_{0}^{r_h}\vert v (r) \vert^2 r^2 dr }\,  .
\end{equation}
In order to find the momentum space version we now define an energy density in momentum space:
\begin{equation} 
 {\tilde v} (k) = \frac{ 1 }{ (2 \pi )^{3/2} } \int v (r)  \exp ( -i \vec k \cdot \vec r ) d^3r     \,.
\end{equation}
After taking the limit $R \to \infty $ one finds
\begin{eqnarray} 
 {\tilde v} (k) &=& \frac{1 }{ (2 \pi )^{3/2} } \frac{3}{2 b^2 G_N} \left( \frac{6 r_h^5 + 4 r_h^3 b^2 }{3 r_h^2 - b^2} \right) \cr & &  \times \left( \frac{ \sin (k r_h ) }{ k^3 }  - \frac{ r_h \cos (k r_h ) }{ k^2 }  \right)  \,,
\end{eqnarray}
that comes only from contributions of the density inside the horizon. 

The modal fraction in momentum space is defined as 
\begin{eqnarray} 
{\tilde \rho (k) } &=& \frac{ \vert {\tilde v}(k) \vert^2 }{ 4 \pi \int_{0}^{\infty }\vert {\tilde v } (k) \vert^2 k^2 dk } \cr &=&   \frac{1}{4 \pi (\pi r_h^3/6)} \,  \left( \frac{ \sin (k r_h ) }{ k^3 }  - \frac{ r_h \cos (k r_h ) }{ k^2 }  \right)^2
\end{eqnarray}
The information entropies in position and momentum spaces are given by eqs. (\ref{positionentropy}) and (\ref{momentumentropy}) , respectively, with spherical symmetry
 \begin{eqnarray}
 f  &=&  - \int d^3r \, \rho({ r}) \log \rho({ r}) \,  ,\cr
  {\tilde f } &=& - \int d^3k \, { \tilde \rho} ({ k}) \log  {\tilde \rho} ({ k}) \,.
 \label{entropiestoghether}
 \end{eqnarray}
 Using eq. (\ref{densities}) one gets: 
 $$ f  =    \log\left(4 \pi r_h^3 /3\right) $$
 
It is important to remark that both $f $ and $ {\tilde f}$ receive contributions from the energy density (\ref{densities}) only from the region inside the horizon:  $ r < r_h$.  This means that the changes in information, as represented by the variations of  $f$ and  ${\tilde f }$,   come only from inside the black hole, as one would expect from physical grounds. There is a non trivial fact to be noticed:  the black hole geometry (\ref{AdSBH}) is defined only for $ r \ge r_h$ but the subtraction of the thermal AdS background (\ref{AdS}) leads to the appearance of an energy density inside the horizon. And this is precisely the density that contributes to the  information entropies.

\section{Results and Conclusions}  
 We show in figure {\bf 1}  plots of the  entropies $f$, ${ \tilde f} $ and their sum:  $f + { \tilde f} $ as a function of the horizon radius divided by the AdS radius $r_h/b$. One notes that as the horizon radius, and correspondingly the temperature, increase, the   entropy in position space $f$ increases while the momentum space entropy ${ \tilde f } $ has the opposite behavior. 
 They satisfies the constraint given by the uncertainty relation in eq. (\ref{uncertainty}) and actually behave in a trivial way: the sum is  constant.

\begin{figure}[h]
\label{g67}
\begin{center}
\includegraphics[scale=0.5]{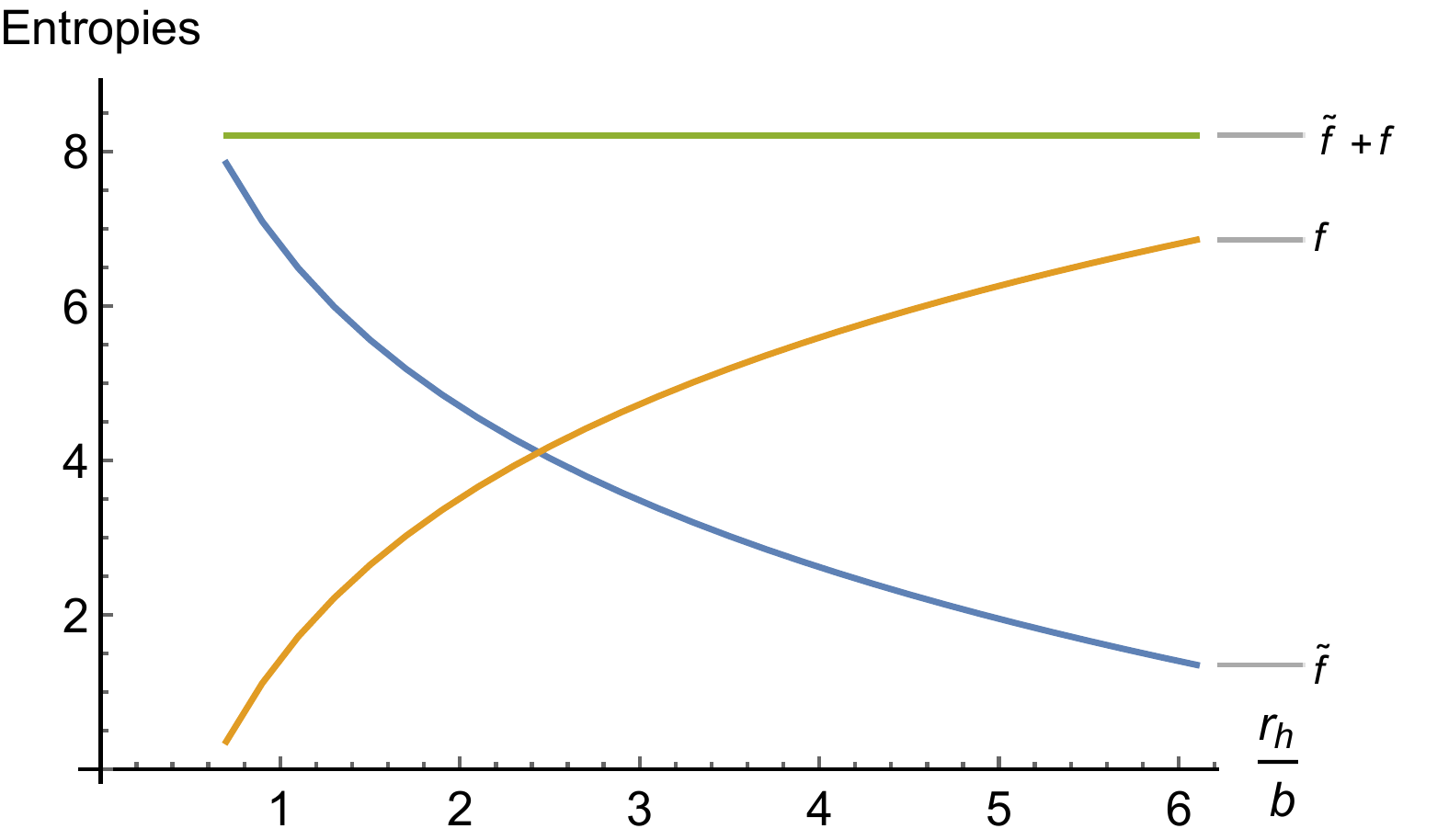}
\end{center}
\caption{ Information Entropies for the AdS black hole as a function of $r_h/b$.  Blue line: momentum space entropy. Orange line: position space case: $f$ . Green line: sum of the entropies. }
\end{figure}
 So, as the black hole temperature increases, and the black hole increases the stability against a Hawking Page transition, information encoded in the spatial  distribution of energy increases while information in the momentum distribution decreases.  And the total information associated with the quantities $f$ and $\tilde f$ remains constant. 
 
 It is not clear, at this moment,  if such a behavior is particular of the AdS black hole or could be more general. Previous studies of configuration entropy, like 
 \cite{Gleiser:2013mga,Gleiser:2015rwa,Casadio:2016aum} indicate that the momentum space entropy decreases with stability, but for the spatial entropy, and their sum, we are not aware of any similar study. This could be an interesting topic for future investigation. 
 
 Regarding stability,   the black hole states are  dominant for temperatures above the critical one, that corresponds to $ r _h/b = 1$.   The semi-classical argument of  Hawking and Page is that the relative probability of a given configuration is proportional to the exponential of (minus) the corresponding action.  So the probability of finding the black hole is governed by the factor $\exp \left(\, - I  \, \right)$, where $I$ is the difference between the black hole action and the thermal action.  
The larger the ratio $r_h/b$, the larger  the difference between the actions.  
So that, the  black hole dominance increases smoothly  with  $r _h/b$.  This is consistent with the decrease in the configuration entropy $ { \tilde f }$ found here, that is shown in  figure {\bf 1} . 
There is  no particular signature of the  Hawking-Page transition point $ r _h/b = 1$ from the point of view of the configuration entropy. This can be explained by the fact that the variation in the difference between 
 the  black hole and  thermal AdS actions with the temperature is smooth.    
 
It is interesting to note that if one looks at the logarithms of the entropies, one finds for $  { \tilde f }$  a simple scaling relation. We plot  $ \log [ { \tilde f }  (r_h/b) ] $ in figure {\bf 2}. There is an approximate linear behavior of the form: $ \log ({\tilde f})  = 2.20 - 0.31 (r_h/b) $.  For the logarithm of the position entropy there is no such linear fit.

\begin{figure}[h]
\label{g67}
\begin{center}
\includegraphics[scale=0.5]{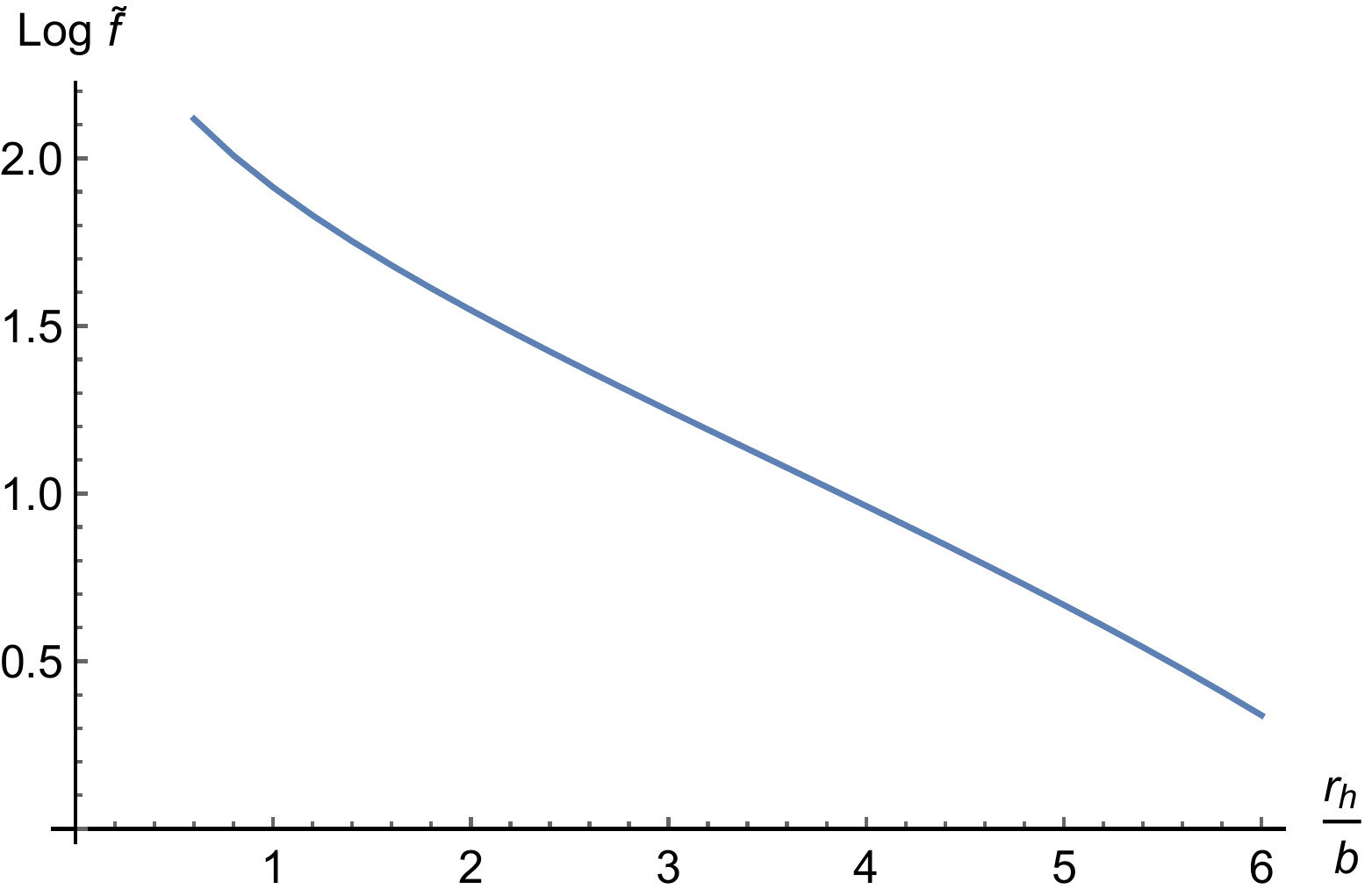}
\end{center}
\caption{ Logarithm of the momentum space entropy  }
\end{figure}

\noindent {\bf Acknowledgments:}    The authors thanks  Alfonso Ballon-Bayona,  Luiz Davidovich and Nicin Zagury
for important discussions. This work is partially supported by  CNPq    (grant  307641/2015-5) and by Coordenação de
Aperfeiçoamento de Pessoal de Nível Superior - Brasil (CAPES) - Finance Code 001.

 \end{document}